\documentclass{aastex6}

\begin{document}

\title{Extra-galactic distances with massive stars: the role of stellar variability in the case of M33}

\author{Chien-Hsiu Lee\altaffilmark{1}}
\affil{Subaru Telescope, NAOJ, 650 N Aohoku Pl, Hilo, HI 96720, USA}

\altaffiltext{1}{leech@naoj.org}

\begin{abstract}
  In modern cosmology, determining the Hubble constant (H$_0$) using distance ladder to percent level and comparing with the results
  from Planck satellite can shed light on the nature of the dark energy, the physics of neutrino, and the curvature of
  the universe. Thanks to the endeavor of the SH0ES team, the uncertainty of the H$_0$ has be dramatically reduced from
  10\% to 2.4\%, with the promise to reach even 1\% in the near future. In this regard, it is fundamentally important to beat and investigate the systematics. This is best be done with other independent good distance indicators.
One of the promising method is the flux-weighted gravity luminosity relation (FGLR) of the blue supergiants. As the blue supergiants are the brightest
  objects in the galaxies, they can probe distance up to 10 Mpc, with negligible blending effects. While the FGLR method delivered distance
  in good agreement with other distance indicators, it has been shown that this method delivers larger distance in the case of M33 and NGC 55.
  Here we investigate whether the M33 distance estimate of FGLR suffers systematics from stellar variability. Using CFHT M33 monitoring data,
  we found 9 out of 22 BSGs showed variability during the course of 500 days, however with amplitudes as small as 0.05 magnitudes. This suggests
  that stellar variability plays negligible role in the FGLR distance determination.
\end{abstract}

\keywords{Galaxies: distances and redshifts -- galaxies: individual (M33) -- galaxies: individual (NGC 55) -- stars:early-type -- supergiants}

\section{Introduction}
Determining extra-galactic distance with exquisite precision and narrowing down 
Hubble constant (H$_0$) to 1-2\% have been a long quest for modern observational astronomy.
Accurate and precise H$_0$ measurement from the distance ladder method, when compared with the
cosmic microwave background (CMB) results from the Planck satellite, can provide constraints 
on the equation of state of dark energy, the mass of neutrino, and the spatial curvature of the
unviverse \citep{2005ASPC..339..215H, 2016A&A...594A..14P}. We note that to extract H$_0$ from the CMB measurements requires an assumption about the cosmology of our universe, for instance that the universe is flat. A discrepancy between H$_0$ obtained from CMB measurements and other direct local H$_0$ measurements indicates that the applied standard cosmology is not correct.

Thanks to the endeavor of the SH0ES team \citep{2016ApJ...826...56R}, the 
H$_0$ uncertainty has been dramatically improved from 10\% to 2.4\%, with the promise of 1\% in the near future. In this regard, understand the systematic errors associated with the present established methods is necessary. If we want to reach 1\% precision with H$_0$, then beating and investigating the systematics is fundamentally important and this is best be done with other independent good distance indicators.

Under this context, independent distance estimates other than the Cepheid method are 
pivotal. One promising method is to use the blue supergiants (BSG), as proposed by 
\cite{2003ApJ...582L..83K,2008ApJ...681..269K}. While the FGLR method delivers distances in basic agreement with other
distance indicators in several extra-galactic systems, there are some marginal
discrepancies in M33 and NGC55, both toward a larger 
distance from FGLR. In this work we investigate whether such discrepancies could
  originate from stellar variability.

Our paper is organized as follows. We provide an overview of the FGLR method
  and its tension with Cepheids distance estaimte in the case of M33 and NGC 55 in section \ref{sec.tension}. 
In section \ref{sec.obs} we investigate the stellar variability of M33 BSGs using CFHT observations,
followed by a discussion and summary in section \ref{sec.sum}.

\section{FGLR distance estimate, and its tension with Cepheids distance in the case of M33 and NGC 55}
\label{sec.tension}

The BSG is a short phase when massive stars (12 to 40 M$_\odot$)
evolve off main-sequence and towards the red supergiant. At this stage, the mass and luminosity
are roughly constant \cite{2000A&A...361..101M}, meaning the surface gravity ($g$) and 
effective temperature (T$_{eff}$) are coupled as $g$/T$_{eff}^4$ = constant. \cite{2003ApJ...582L..83K,2008ApJ...681..269K} then defined flux-weighted gravity g$_F$ = $g$/T$_{eff}^4$. Assuming
the usual mass-luminosity, L $\propto$ M$^\alpha$, where $\alpha\sim$ 3, \cite{2003ApJ...582L..83K,2008ApJ...681..269K} thus derived a relation with the absolute bolometric magnitude M$_{bol}$
and the flux-weighted gravity as:
\begin{equation}
M_{bol} = a (\mathrm{log}g_F - 1.5) + b,       
\end{equation}

where a=3.41 and b=-8.02 according to \cite{2008ApJ...681..269K}. 

The advantage of the BSG method is multi-fold. As the brightest star in the optical wavelength,
they can be used to determine distance up to 10 Mpc. Their bright nature also implies that they
are hardly affected by the blending effects (which is often a concern for Cepheids). 
In addition, from broad-band photometry, their line-of-sight extinction can be well constrained
(which is an issue for Cepheids PL with optical photometry).   
The BSG method has been applied to WLM \citep{2008ApJ...684..118U}, M33 \citep{2009ApJ...704.1120U},
M81 \citep{2012ApJ...747...15K}, M106 \citep{2013ApJ...779L..20K}, 
NGC3621 \citep{2014ApJ...788...56K}, NGC3109 \citep{2014ApJ...785..151H}, and NGC55 \citep{2016ApJ...829...70K}.

We then compare the FGLR derived distances with other methods, as shown in Table \ref{tab.distance}. For illustrational
purpose, we use the mean and standard deviation of all distance estimates in the literature, extracted from the 
NASA Extra-galactic Database\footnote{https://ned.ipac.caltech.edu}. While most of the distance estimates from FGLR are 
in good agreement with other distance indicators, there are two systems, i.e. M33 and NGC 55, both show 
farther distance from FGLR than other distance indicators. This suggests that either other distance indicators all
deliver shorter distances, or the BSGs used in FGLR are fainter than expected. We recall that the FGLR methods 
take into account the line-of-sight dust extinction, and the BSGs are so bright that they are hardly affected by
blending, thus we can rule out the contaminations from dust extinction or blending. To have a fair comparison,
we now consider other distance indicators with high precision. Here we focus to the near-infrared Cepheid PL 
relation, a very reliable technique which is also cited by the FGLR working group \citep{2016ApJ...829...70K}. 
Though Cepheid PL in optical may suffer from effects such as dust extinction and metallicity, these effects are
negligible in the infrared. Furthermore, with exquisite angular resolution from HST or ground-based AO, we can 
correct for the blending effects.

\begin{table}
\centering
\caption{Distance modulus from FGLR and other methods}
\begin{tabular}[t]{lrlr}
  \hline
  \hline
Name & FGLR & Cepheids IR PL$^\dagger$ & NED $^\ddagger$ \\
     & [mag] & [mag] \\
\hline
WLM & 24.99$\pm$0.10 & 24.924$\pm$0.042(statistic)$\pm$0.065(systematic)$^1$ & 25.00$\pm$0.47 \\
& & & \\
M33 & 24.93$\pm$0.11 & 24.62$\pm$0.07$^2$ & 24.68$\pm$0.34 \\
& & & \\
M81 & 27.70$\pm$0.10 & -- & 27.82$\pm$0.32 \\
& & & \\
NGC3621 & 29.07$\pm$0.09 & -- & 29.16$\pm$0.27 \\
& & & \\
NGC3109 & 25.55$\pm$0.09 & 25.571$\pm$0.024(statistic)$\pm$0.065(systematic)$^3$ & 25.60$\pm$0.29 \\
& & & \\
NGC55 & 26.85$\pm$0.10 & 26.434$\pm$0.037(statistic)$\pm$0.087(systematic)$^4$ & 26.41$\pm$0.32 \\
\hline
\hline
\multicolumn{4}{l}{$\dagger$ The values are taken from the Araucaria project, which delivers both FGLR and }\\ 
\multicolumn{4}{l}{Cepheid IR PL distance estimates. As both the FGLR and Cepheid IR PL distances}\\
\multicolumn{4}{l}{are from the same working group, this provides a high accuracy, consistency check}\\
\multicolumn{4}{l}{of the FGLR method. 1) from Gieren et al. (2008); 2) from Gieren et al. (2013);}\\
\multicolumn{4}{l}{3) from Soszynski et al. (2006); 4) from Gieren et al. (2008).}\\
\multicolumn{4}{l}{$\ddagger$ The values are taken from the NASA Extra-galactic Database, using the mean}\\ 
\multicolumn{4}{l}{and standard variation from all the distance estimates in the literature.}
\end{tabular}
\label{tab.distance}
\end{table}
        
The Araucaria project has applied the NIR Cepheid PL and obtained an M33 distance modulus of 24.62$\pm$0.07 mag
\citep{2013ApJ...773...69G}, in line with the vast majority of the distance anchor methods with better precision, 
but differs significantly (0.31 mag shorter) with the FGLR method. As for the case of NGC 55, the Araucaria 
project obtained an NIR Cepheid distance modulus of 26.43$\pm$0.09 mag \citep{2008ApJ...672..266G}, which is 0.42 
mag smaller than the FGLR distance estimate. While \cite{2016ApJ...829...70K} speculated the differences between 
NIR Cepheids and FGLR distances of NGC 55 to stem from the blending effects, especially because NGC 55 is an 
edge-on galaxy, this is not the case for M33, which is a face-on galaxy. As the NIR Cepheid PL is a very reliable
method, this calls for further investigations of the FGLR method, especially the variability of the 
BSGs employed by the FGLR method.

\section{Variability survey of M33}
\label{sec.obs}

We notice that there were works on variability influences on FGLR with NGC 300 and NGC 55. For NGC 300, \cite{2004ApJ...600..182B} have shown that BSG variability while present with an amplitude of 0.05 mag or smaller has a negligible effect on FGLR distances. In addition, \cite{2008ApJ...681..269K} use this galaxy as one of their FGLR calibrators, leading to a=3.41 and b=-8.02 based on the Araucaria Cepheid distance. For NGC 55, \cite{2016ApJ...829...70K} use only targets with variability amplitude smaller than 0.05 mag -- based on the comprehensive study by \cite{2012A&A...542A..79C} -- and include this amount of variability in their distance error estimate. In addition to NGC 55 and NGC 300, M33 is the galaxy under scrutiny in this paper. Because of the discrepancy it is an important case to investigate the effects of variability again.

Luckily, there was a deep M33 variability survey conducted by the 
CFHT telescope \citep{2006MNRAS.371.1405H}. \cite{2006MNRAS.371.1405H} made use of the MegaCAM on-board CFHT; with MegaCAM's
wide field-of-view (1x1 deg$^2$), the entire M33 galaxy can be observed with one single shot. This survey was carried out
in three seasons between 2003-2005, with 3 Sloan filters gri, and obtained a total of 33 epochs. As M33 is a very crowded
stellar field, the data were analyzed using image subtraction process proposed by \cite{1998ApJ...503..325A}. This allows
to extract high precision photometry down to the limiting magnitude even in the very crowded stellar field like M33.
\cite{2006MNRAS.371.1405H} then go through the resolved stars in the M33 images, and detect 36709 varying sources that
show variability at $>$ 5 $\sigma$ level compared to a constant flux light curve.

\begin{figure}
\includegraphics[scale=0.7]{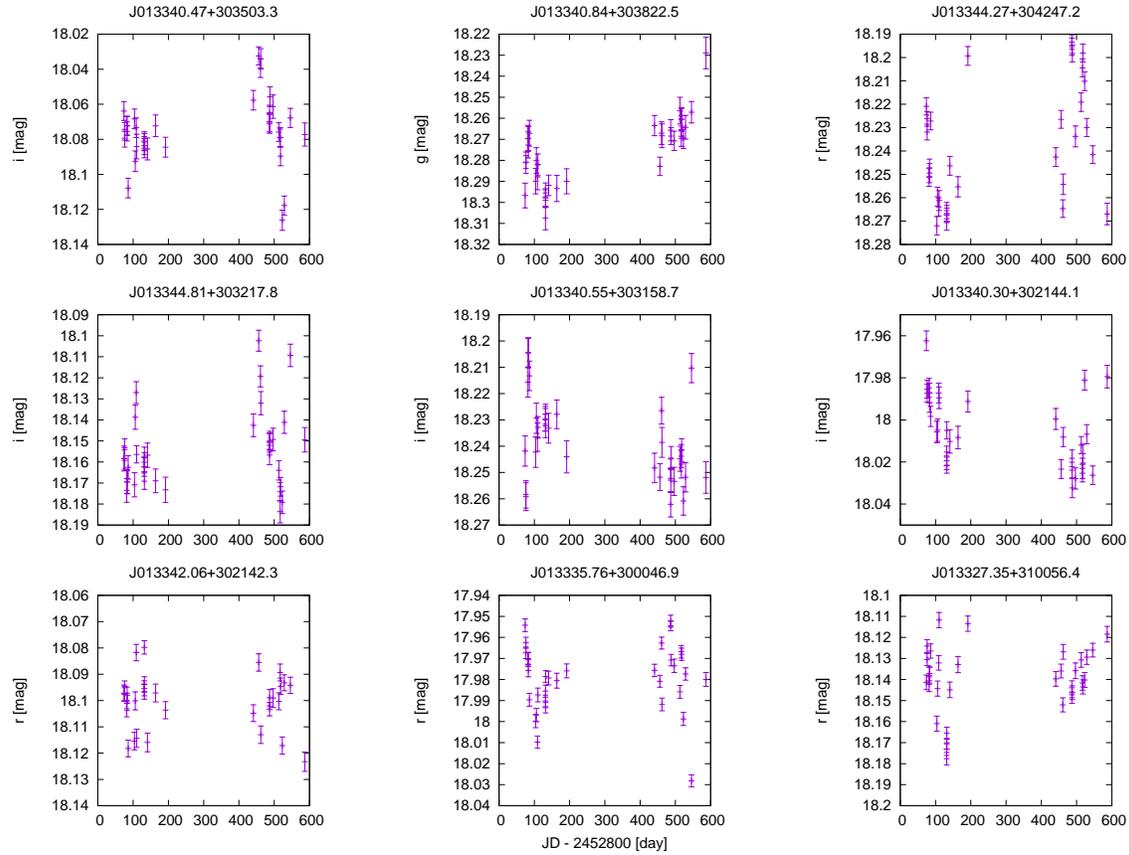}
\caption{Light curves of 9 M33 variable BSGs used in the FGLR distance estimate \citep{2009ApJ...704.1120U}. Data gathered from CFHT M33 variability \citep{2006MNRAS.371.1405H} in Sloan g, r, and i-band.}
\label{fig.cfht}
\end{figure}

We then search the variable catalogue from \cite{2006MNRAS.371.1405H} for the 22 BSGs that were used in the \cite{2009ApJ...704.1120U}
FGLR work, and found that 9 of them show variability from the 33 epochs of the CFHT data, up to 0.1 mag level variation, or 0.05 mag in semi-amplitude,
as shown in Fig. \ref{fig.cfht}.

\section{Discussion and Summary}
\label{sec.sum}

As has been shown by the CFHT multi-epoch data, the M33 BSGs used in \cite{2009ApJ...704.1120U} exhibit the level of variability
  in the same order of magnitudes as seen in BSGs in NGC 300 \citep{2004ApJ...600..182B}. As has been discussed in \citep{2004ApJ...600..182B}, this
  has negligible effect on the FGLR distance estimates. Even more so, such variability has been taken into account in the uncertainties of the
  FGLR distance determination (see e.g. Kudritzki et al. 2016), which is an empirically calibrated relation using large sample of stars (see e.g.
  Kudritzki et al. 2003, 2008, Urbaneja et al. 2017).

While there is marginal discrepancy between the FGLR and Cepheid distance estimates of M33, our investigation indicates we can rule out stellar
  variability as the cause of such tension. We note that the distance estimate using eclipsing binaries (Bonanos et al. 2006) also showed comparatively
  larger distances in the case of M33, in agreement with the FGLR method. As we eliminate stellar variability as possible causes of tension
  between FGLR and Cepheid distances, the cause of discrepancies remains an open question, and awaits more lines-of-thoughts and further investigations.

 \acknowledgments
 We are grateful to the referee for the insightful comments which greatly improved this manuscript. The authors wish to recognize and acknowledge the very significant cultural role and reverence that the summit of Maunakea has always had within the indigenous Hawaiian community. We are most fortunate to have the opportunity to make use of observations from this mountain. 
  


\end{document}